\documentclass[aps,pra,10pt,notitlepage,twocolumn,showpacs]{revtex4-1}
\usepackage[dvips]{graphicx}
\usepackage{amssymb,amsfonts}
\usepackage{amsmath}
\newcommand{\bra}[1]    {\langle #1|}
\newcommand{\ket}[1]    {| #1 \rangle}

\newcommand{\opi}[2]{\frac{#2}{(2\pi)^#1}}

\newcommand{\f}[1]{\textrm{#1}}
\newcommand{\B}[1]{\mathbf{#1}}
\newcommand{\x}{{\mathbf r}}
\newcommand{\kk}{{\mathbf k}}

\begin{document}

\author{T. Wasak$^1$, J. Chwede\'nczuk$^1$, P. Zi\'n$^{2,3}$ and M. Trippenbach$^1$}
\affiliation{$^1$Faculty of Physics, University of Warsaw, ul.\ Ho\.{z}a 69, 00-681 Warszawa, Poland\\ 
  $^2$Narodowe Centrum Bada\'n J\c{a}drowych, ul. Ho\.za 69, 00-681 Warszawa, Poland\\
  $^3$Univ. Paris Sud, CNRS, LPTMS, UMR 8626, Orsay 91405 France}

\title{Raman scattering of atoms from a quasi-condensate in a perturbative regime}

\begin{abstract}
  It is demonstrated that measurements of positions of atoms scattered from a quasi-condensate in a Raman process provide information on the temperature of the parent cloud.
  In particular, the widths of the density and second order correlation functions are sensitive to the  phase fluctuations induced by non-zero temperature of the quasi-condensate.
  It is also shown how these widths evolve during expansion of the cloud of scattered atoms. 
  These results are useful for planning future Raman scattering experiments and indicate the degree 
  of spatial resolution of atom-position measurements necessary to detect the temperature dependence of the quasi-condensate.
\end{abstract}


\maketitle

\section{Introduction}

Atoms scattered out of Bose-Einstein condensates can be an object of benchmark tests of various quantum-mechanical models. 
A prominent example is a collision of two counter-propagating condensates \cite{kozuma,vogels,paris1,paris2,paris3,australia,phil}. 
During the collision, which takes place at super-sonic velocity, atoms are scattered into initially empty modes, and description of such
process requires fully quantum treatment. This can be done semi-analytically in the Bogoliubov approximation  \cite{bach,chwed1,chwed2,chwed3,chwed4} or numerically in more general cases
\cite{ballagh,deuar1,deuar2,deuar3,deuar4}. The analysis reveals strong correlations between the scattered atoms \cite{paris1,chwed1,chwed2,chwed3,chwed4} 
and sub-poissonian fluctuations of the opposite-momentum atom counts \cite{paris3}. Therefore, the many-body atomic states created in the collisions could have potential application
for ultra-precise sub shot-noise atomic interferometry \cite{schmied}. 

A different relevant example of atom scattering out of a coherent cloud takes place in a spin-1 condensate \cite{klempt1,klempt2,hyllus}. In this case, a single
stationary matter-wave is prepared in a Zeeman substate with $m_F=0$. A two-body interaction can change the spin projection of the colliding pair into $m_F=\pm1$. Recently, it has been
demonstrated \cite{hyllus} that produced atomic pairs are usefully entangled from atom-interferometry point of view.

Here we concentrate on another pair production process, namely the Raman scattering \cite{polzik,simon}. 
In this case, an ultra-cold atomic cloud is  illuminated with a strong laser beam. As a
result, an inter-atomic transition leads to creation of a correlated Stokes photon and atomic excitation. The scattered pairs are correlated analogously 
to those produced in the condensate- or spin-changing collisions. Raman scattering is similar to the elastic Rayleigh process 
\cite{ketterle1,ketterle2}, though the Stokes photons have different energy then the incident light. This process has been widely studied theoretically 
\cite{meystre1,wang,meystre2,deng, zhou} and observed experimentally in ultra-cold samples 
\cite{yoshikava2} and Bose-Einstein condensates \cite{ketterle3,yoshikava1,yoshikava3,sadler}.
In this work we consider a different source of Raman-scattered particles, 
namely the quasi-condensate, which forms in strongly elongated traps \cite{shapiro2,shapiro1,shapiroReview}. 
Due to non-zero temperature of the gas, phase fluctuations occur and they limit the spatial coherence of the system. This, in turn, has influence on the scattering process. 
We demonstrate that one can determine the temperature of the parent cloud
from both the density and the second order correlation function of the scattered atoms.
\begin{figure}[htb]
  \centering
  \includegraphics[scale=0.44,clip]{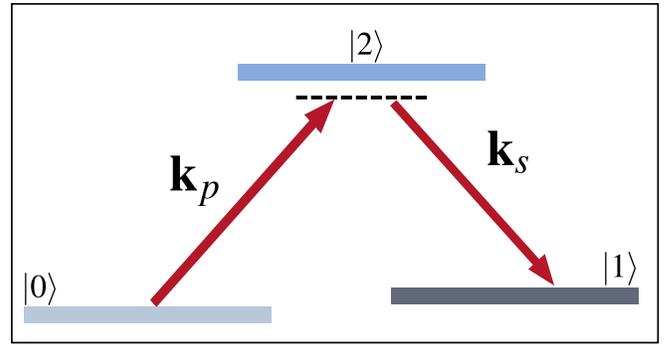}
  \caption{Scheme of atomic levels relevant in the process of Raman scattering. The atom, initially in state $|0\rangle$, absorbs strongly detuned pump photon with wave-vector
    $\kk_p$. The absorption is accompanied by spontaneous emission of a Stokes photon with wave-vector $\kk_s$. As a result, the atom undergoes a transition $0\rightarrow1$.}
  \label{fig:levels}
\end{figure}

The paper is organized as follows. 
In Section II we formulate the 3-dimensional problem and introduce the Hamiltonian for the process of Raman scattering. We derive the Heisenberg equations for atoms and photons
and introduce the relevant correlation functions. 
In Section III, basing on perturbative solution of the atomic dynamics, we calculate the one-body density matrix both in the position and momentum representation. 
In Section IV we discuss the method for incorporating the phase fluctuations due to non-zero temperature of the quasi-condensate. Using this approach, 
we calculate the density of scattered atoms both after long expansion time, i.e. in the far-field regime, and when the expansion time is short. 
Then, we turn to the second order correlation function. We show, how the temperature influences its peak height as well as the width. Some
details of calculations are presented in the Appendices.

\section{Formulation of the model}

The process of Raman scattering takes places when an atom in a three-level Lambda configuration is illuminated with an intense pump beam. 
As a result of interaction with light, the atom absorbs a photon from the pump and undergoes an effective transition $0\rightarrow2\rightarrow1$ accompanied by 
spontaneous emission of a ``Stokes'' photon shown in Fig.~\ref{fig:levels}. 

To model the phenomenon, we assume that the pump can be described classically as
\begin{equation}
  E_p(\x,t)=E_0(\x,t)e^{i(\kk_p\cdot\x - \omega_p t)}+\mathrm{c.c.},
\end{equation}
where $E_0$ is its amplitude, $\kk_p$ is the central wave-vector and $\omega_p=c|\kk_p|$. When this frequency is strongly detuned from the transition $0\rightarrow2$, the upper level
can be adiabatically eliminated. As a result, the process can be regarded as creation of a quantum of atomic excitation $0\rightarrow1$ together with an emission of a Stokes photon.

We describe the quantum state of the atoms and Stokes photons using two field operators
\begin{eqnarray}
    \hat b(\kk,t)&=&\frac1{\sqrt N}\sum_\alpha e^{i(\kk \cdot \x_\alpha - \omega_{01}t)} \ket0_\alpha\bra1_\alpha \label{def_b},\\
    \hat E_S^{(+)}(\x,t)&=&\int\!\!d\kk\ e^{i(\kk\cdot \x - c|\kk|t)} \hat a(\kk,t),
\end{eqnarray}
where $\omega_{01}$ is the $0\rightarrow1$ transition frequency.
The operator $\hat b(\kk,t)$ annihilates an atomic excitation with momentum $\hbar\kk$, while the index $\alpha$ runs over all the $N$ atoms in the cloud. 
If the majority of atoms occupy $\ket 0$, one can apply  the Holstein-Primakoff approximation  \cite{holst}, 
and accordingly  $\hat b$ satisfies bosonic commutation relations. Moreover, $\hat E_S^{(+)}$ is the field operator of the Stokes photons.

When  a large number of atoms $N$ occupy a single-particle state, 
one can replace summation over separate particles in Eq.(\ref{def_b}) with an integral over the quasi-condensate wave-function $\psi(\x,t)$. 
The effective Hamiltonian for the process of Raman scattering is $\hat H=\hat H_0+\hat H_{\rm int}$, where
\begin{equation}\label{ham}
  \hat H_0 = \int\!\!d\kk\,\hbar \omega_k \hat a^\dagger(\kk) \hat a(\kk) + \int\!\! d\kk\, \epsilon_k\hat b^\dagger(\kk) \hat b(\kk)
\end{equation}
is the free part, with $\epsilon_k=\hbar^2 \kk^2/2m$. Also, 
$\omega_k=c|\kk|-\omega_S$ is centered around the Stokes frequency $\omega_S=\omega_p-\omega_{01}$. 

The interaction Hamiltonian
\begin{equation}\label{hint}
  \hat H_{\rm int}=\int\!\!d\kk d\kk' h(\kk,\kk')\ \hat a^\dagger(\kk,t)\hat b^\dagger(\kk',t)+\mathrm{H.c.},
\end{equation}
governs the desired process, where an atomic excitation is created together with the Stokes photon. 
The coupling function $ h(\kk,\kk')$ is expressed in terms of a Fourier transform of the product of the quasi-condensate and pump beam fields,
\begin{equation}
  \label{h_def}
  h(\kk,\kk') = h_0 \int\!\! d\x e^{-i(\kk + \kk' - \kk_p)\cdot \x} \psi(\x,t) E_0(\x,t),
\end{equation}
\begin{figure}[htb]
  \centering
  \includegraphics[scale=0.44,clip]{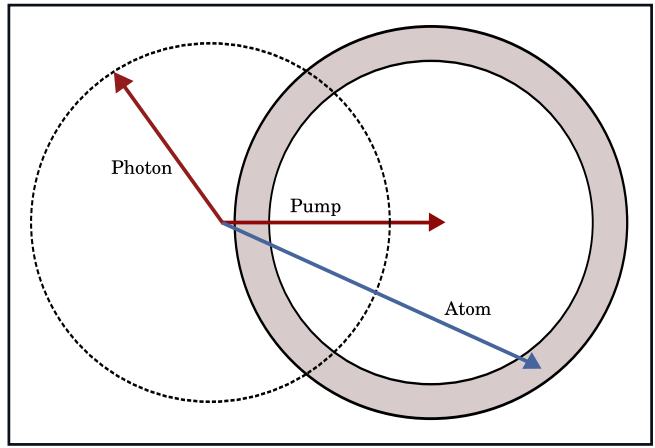}
  \caption{Scheme of the Raman scattering in the momentum space. Spontaneously emitted Stokes photon acquires momentum $k_s$. After many scattering events, the photons will form a 
    sphere of radius $k_s$ denoted here by the dashed circle.
    Due to momentum conservation, atoms scatter onto a sphere of radius $k_s$ as well, shifted by $k_p$ due to absorption of the pump photon. The width of the gray 
    ring occupied by the atoms
    represents the uncertainty resulting from the momentum spread of the parent quasi-condensate.}
  \label{fig:schematic_halo}
\end{figure}
where the coupling constant \cite{most} is equal to
\begin{equation}\label{h0}
  h_0=\sqrt{\frac{\omega_s}{\hbar^{3/2}\varepsilon_0}}\frac{d_{02}d_{21}}{(2\pi)^3}\frac{2\omega_{02}}{\omega_{02}^2-\omega_s^2}. 
\end{equation}
Here, $d_{ij}$ is the atomic dipole moment associated with the $i\rightarrow j$ transition and $\varepsilon_0$ is the dielectric constant. Note that in Eq.~(\ref{hint}) we
have neglected the interaction of the scattered atoms with the mean-field of the quasi-condensate.

Below we make further, physically well justified simplifications. First, we choose the pump envelope $E_0$ to be time-independent, which corresponds to a common situation of 
square pulses. Moreover, the spatial extent of the pump usually vastly exceeds the size of the quasi-condensate. Since the duration of the pump pulse is much shorter than 
the characteristic time scale of $\psi(\x,t)$ dynamics, the quasi-condensate wave-function can be taken constant and the coupling function in a frame of reference moving 
with velocity $\hbar\kk_p/m$ reads
\begin{equation}
  \label{h_def_fourier}
  h(\kk,\kk') = h_0 E_0 \tilde\psi(\kk + \kk').
\end{equation}
Here, $\tilde\psi$ is a Fourier transform of wave-function $\psi$. 

We can now derive the set of coupled Heisenberg equations of motion for the Stokes and atomic field resulting from the Hamiltonian (\ref{ham}),
\begin{subequations}\label{heis}
  \begin{eqnarray}
    i\hbar \partial_t \hat a(\kk,t) &=& \hbar \omega_k \hat a(\kk,t) + \int\!\!d\kk' h(\kk,\kk') \hat b^\dagger(\kk',t) ,\label{heis_a}\\
    i\hbar \partial_t \hat b(\kk,t) &=& \epsilon_k \hat b(\kk,t) + \int\!\!d\kk' h(\kk',\kk) \hat a^\dagger(\kk',t).\label{heis_b}
  \end{eqnarray}
\end{subequations}

These equations are a starting point for the analysis of the second order correlation function of scattered atoms, defined as
\begin{equation}
  G^{(2)}(\kk_1,\kk_2, t)=\langle\hat b^\dagger(\kk_1,t)\hat b^\dagger(\kk_2,t)\hat b(\kk_2,t)\hat b(\kk_1,t)\rangle.
\end{equation}
Since Equations (\ref{heis}) are linear and the initial state of scattered atoms and photons is a vacuum, then
\begin{eqnarray}
  G^{(2)}(\kk_1,\kk_2, t)&=&G^{(1)}(\kk_1,\kk_1,t)G^{(1)}(\kk_2,\kk_2,t)\nonumber\\
  &+&|G^{(1)}(\kk_1,\kk_2,t)|^2,\label{g2}
\end{eqnarray}
is a function of the (one-body) density matrix, which reads
\begin{equation}
  G^{(1)}(\kk_1,\kk_2,t)=\langle\hat b^\dagger(\kk_1,t)\hat b(\kk_2,t)\rangle.\label{g1}
\end{equation}
Its diagonal part $n(\kk,t)=G^{(1)}(\kk,\kk,t)$ represents the momentum distribution of scattered atoms.

In the following section we derive analytical expressions for the density matrix in momentum and position representations 
treating the atom-photon interactions in the perturbative manner. 

\section{Perturbative solutions for atoms}

When the number of scattered atom-photon pairs is small, one can solve Eq.~(\ref{heis_b}) perturbatively in the coupling constant $h_0$ defined in Eq.~(\ref{h0}), 
\begin{equation}\label{pert_dynam_b}
  i\hbar \partial_t \hat b^{(1)}(\kk,t) = \epsilon_k \hat b^{(1)}(\kk,t)+\!\! \int\!\!d\kk' h(\kk',\kk) \hat a^\dagger(\kk')e^{i\omega_{k'}t},
\end{equation}
where $\hat a^\dagger(\kk')=\hat a^\dagger(\kk',0)$. As we argue in Appendix \ref{app_time_dep_b}, since $\epsilon_k\ll\hbar\omega_k$, the first order solution can be written as
\begin{eqnarray}
  &&\hat b^{(1)}(\kk,t) =  e^{-i\frac{\epsilon_k t} \hbar} \hat{b}(\kk)+\label{b_time_dependence}\\
  &&+\frac{t e^{-i\frac{\epsilon_kt}{2\hbar}}}{i\hbar} \int\!\!d\kk' h(\kk,\kk')\hat a^{\dagger}(\kk')\,\f{sinc}\!\left( \frac{\omega_{k'}t}{2}\right)e^{i\frac{\omega_{k'}t}{2}}. \nonumber
\end{eqnarray}
This expression is used to calculate the first order correlation function (\ref{g1})  of scattered atoms.

The measurement of positions of scattered atoms is performed as follows. First, the initial wave-packet of the quasi-condensate interacts with
the pump beam for a time $t_p$ and atoms scatter out of the mother cloud. Then, the system freely expands for a time $t_f$, and finally the positions of atoms are recorded. 
If $t_f$ is sufficiently long and the system reaches the far-field regime, positions of atoms $\x_i$ are related to their wave-vectors $\kk_i$ by 
$\kk_i=\frac{\x_i m}{\hbar t_f}$. 

In the present work we compare the two possible experimental situations, when the system either is or is not in the far field. 
In the former case, as we argued above, it is sufficient to calculate the density matrix (\ref{g1}) in the momentum space just after the interaction ends. 
In the latter, we provide an expression for $G^{(1)}$ as a function of expansion time in position space.

\begin{figure}[tb]
  \centering
  \includegraphics[scale=0.28,clip]{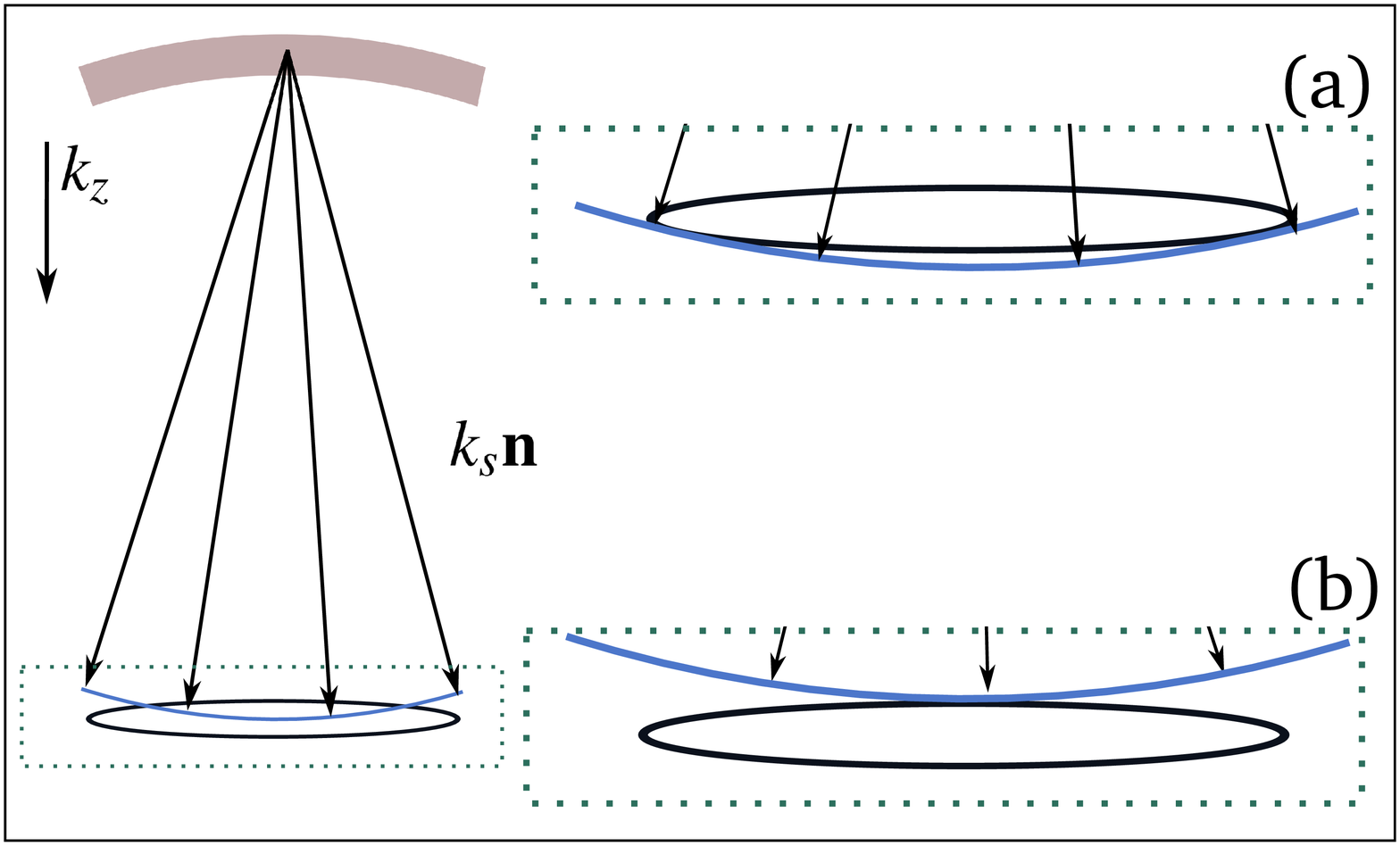}
  \caption{Graphical representation of the density integral (\ref{dens}) for $\kk\simeq k_s{\bf e}_z$. The gray ring represents the halo of scattered atoms, 
    in analogy to Fig.~\ref{fig:schematic_halo}. When $\kk=k_s{\bf e}_z$, as in the main part of the picture,
    the integration samples the majority of the atomic cloud, leading to value of the peak height. (a) When $|\kk|<k_s$, the integration samples the tails of the quasi-condensate and thus
    the density of scattered atoms does not vanish. (b) However, 
    when $|\kk|>k_s$, the tails do not contribute to the integral anymore and the density drops rapidly with growing $k$.}
  \label{den_int}
\end{figure}

\subsection{Momentum-dependent density matrix}

In order to calculate the momentum-dependent density matrix, 
note that for typical interaction times, the ``sinc'' function appearing in Eq.~(\ref{b_time_dependence}), which is peaked around $k'=k_s$, has much smaller width
than the Fourier transform of the condensate function, which, via Eq.~(\ref{h_def_fourier}), enters $h(\kk,\kk')$. Therefore, one can fix the length of the photon wave-vector
to be equal to $k_s$. 
Using the definition from Eq.~(\ref{g1}) and the solution from Eq.~(\ref{b_time_dependence}) we obtain the density matrix in the momentum representation,
\begin{equation}
  G^{(1)}(\kk_1,\kk_2) = \alpha\!\!\int\!\!d\Omega'\ \tilde\psi^\star(\kk_1 + k_s \B{n}') \tilde\psi(\kk_2 + k_s \B{n}'),
  \label{G1_momentum}
\end{equation}
where we omitted an irrelevant phase factor and $\alpha = \frac{2\pi t_p}{c\hbar^2}|h_0|^2 |E_0|^2 k_s^2$. 
The integration is performed over all the directions of the unit vector $\B{n}'$. Since the above perturbative expression, apart from a trivial scaling of $\alpha$ with $t_p$, 
is time-independent, we have skipped the time argument of $G^{(1)}$. 
All the intermediate steps leading to the above solution are presented in Appendix \ref{app_G1}.

By setting $\kk_1 = \kk_2=\kk$ we obtain the density of the scattered atoms
\begin{equation}\label{dens}
  n(\kk) = \alpha \int\!\!d\Omega\ |\tilde\psi(\kk + k_s \B{n})|^2,
\end{equation}
which is directly related to the momentum distribution of the quasi-condensate. Integration over all directions of Stokes photon momentum 
$\B n k_s$ is characteristic for a spontaneous regime, where photons scatter isotropically.
Note also that since the quasi-condensate wave-function from Eq.~(\ref{h_def_fourier}) is expressed in the reference frame moving with the velocity $\hbar\kk_p/m$, in the laboratory frame 
scattered atoms form a halo centered around $\kk_p$ vector , see Fig. \ref{fig:schematic_halo}.

\begin{figure}[htb]
  \centering
  \includegraphics[width=0.450\textwidth,clip]{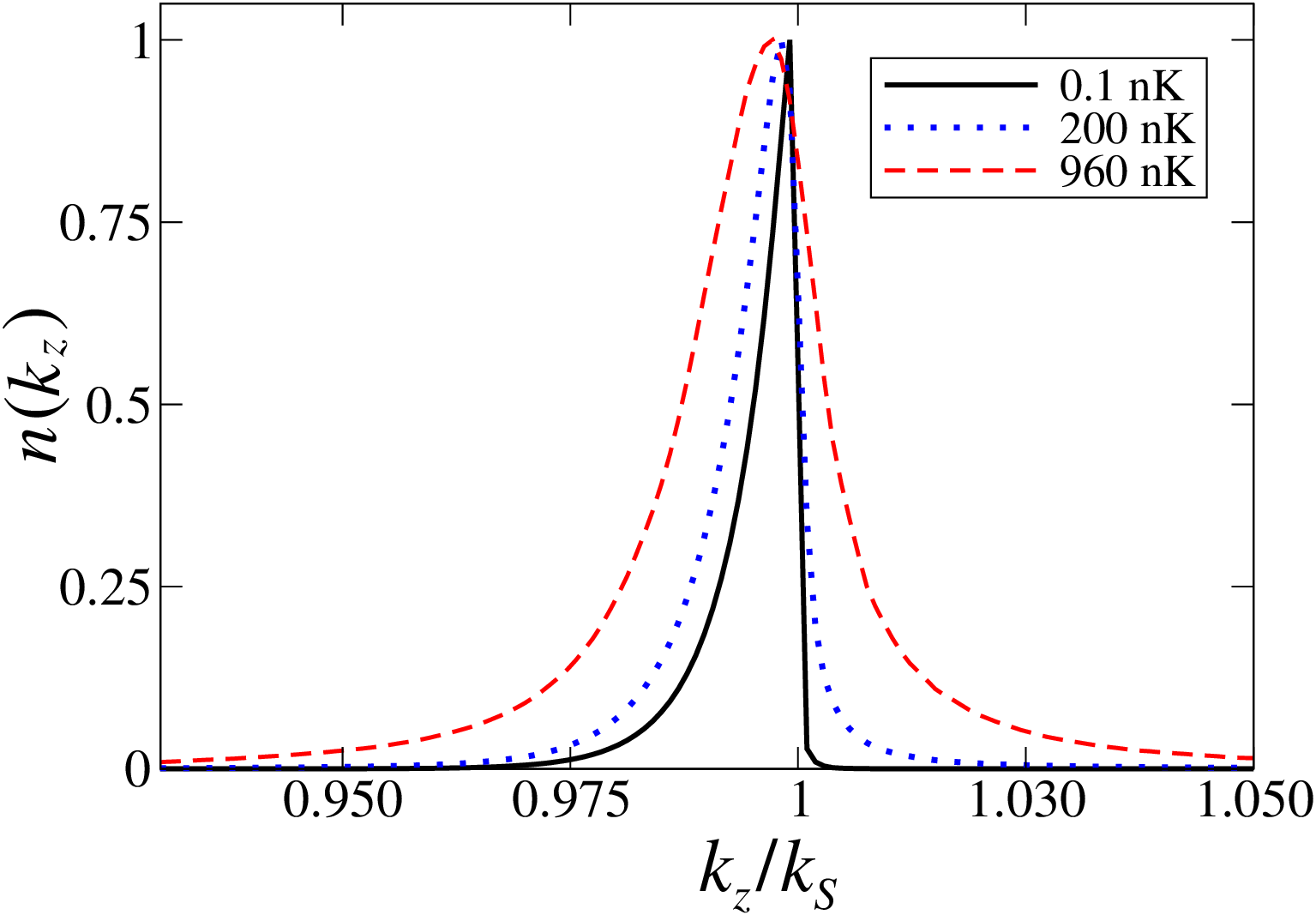}
  \caption{Density of scattered atoms for  $\kk \simeq k_s \B{e_z}$ for different temperatures and normalized by the value of the peak height. 
    Each curve is an average over 400 realizations of the phase noise. The black solid line is calculated for the quasi-condensate at $T=0.1$nK, 
    the dotted blue line for  $T=$200nK and the dashed red line for $T=$960nK.}
  \label{kz_den}
\end{figure}

\subsection{Position-dependent density matrix}
To deal with situations when $t_f$ is not sufficiently long for the system to enter the far-field regime we provide an expression for the density matrix 
in position space as a function of the expansion time $t_f$, which up to an irrelevant phase factor reads
\begin{eqnarray}
  &&\tilde G^{(1)}(\x_1,\x_2,t_f)=\frac{\alpha}{(2\pi)^6}\times\\
  &&\int\!\!d\Omega\,S\left[\tilde\psi\left(\frac{\x_1 m}{\hbar t_f}+k_s \B{n}\right)\right] 
  \cdot S\left[\tilde\psi\left(\frac{\x_2 m}{\hbar t_f}+ k_s \B{n}\right)\right]^\star.\nonumber
  \label{G1_position}
\end{eqnarray}
Here, the functional $S$ is given by
\begin{equation}
  S\left[\tilde\psi(\kk)\right] = \int\!\!d(\delta\kk)\ e^{-i\frac{\hbar (\delta k)^2}{2m}t_f}  \tilde\psi(\kk + \delta\kk ).
\label{def_S}
\end{equation}
The derivation of Eq.~(\ref{G1_position}) is presented in detail in the Appendix \ref{app_position_rep}. Note that (\ref{G1_position}) resembles (\ref{G1_momentum}) 
except that $\tilde\psi$ is replaced with $S[\tilde\psi]$. 
The atomic density is obtained by setting $\x_1=\x_2=\x$ in
Eq.~(\ref{G1_position}) and reads
\begin{equation}\label{dens_pos}
  \tilde n(\x,t_f)=\frac{\alpha}{(2\pi)^6}\int\!\!d\Omega\,\left|S\left[\tilde\psi\left(\frac{\x m}{\hbar t_f}+k_s \B{n}\right)\right]\right|^2.
\end{equation}
We also underline that for sufficiently long $t_f$, $S[\tilde\psi(\kk)]\sim\tilde\psi(\kk )$, so the far field is reached.

\section{Numerical results}

In this section we calculate the correlation functions (\ref{G1_momentum}) and (\ref{G1_position}) using realistic experimental parameters. First, we briefly describe a 
numerical method for simulating phase fluctuations present in a strongly elongated ultra-cold bosonic gas.

\begin{figure}[htb]
  \centering
  \includegraphics[width=0.450\textwidth,clip]{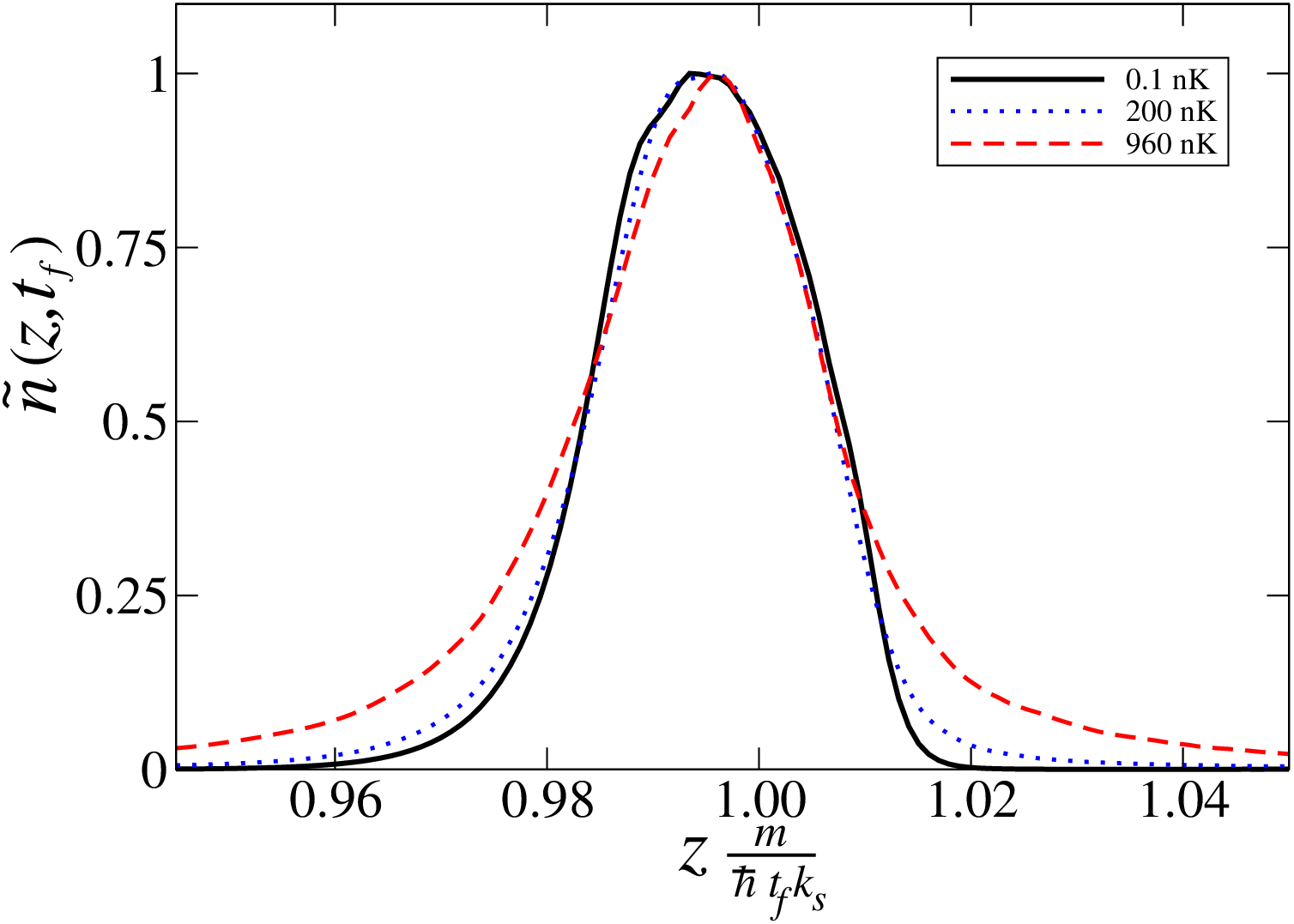}
  \caption{Density of scattered atoms for  $\frac{m\x}{\hbar t_f} \simeq k_s \B{e_z}$ for different temperatures and normalized by the value of the peak height. 
    The time of flight is $t_f=300\,$ms.
    Each curve is an average over 400 realizations of the phase noise. The black solid line is calculated for the quasi-condensate at $T=0.1$nK, 
    the dotted blue line for dashed  $T=$200nK and the dashed red line for $T=$960nK.
  }
  \label{z_den}
\end{figure}

\subsection{Quasi-condensate}

We apply the above model to the process of Raman scattering of atoms from $N=10^5$  metastable $^4$He bosons
with the atomic mass $m = 6.65 \times 10^{-27}\,$kg and the scattering length $a=7.5 \times 10^{-9}\,$m. 
The atoms are confined in a harmonic potential with the radial and the axial frequencies 
equal to $\omega_r = 2\pi\times 1500\, \frac1{\rm s}$, $\omega_z = 2\pi\times 7.5\, \frac1{\rm s}$. 
Such an elongated gas is called a ``quasi-condensate'' due to presence of the phase fluctuations along the $z$ axis \cite{shapiro2,shapiro1,shapiroReview}.

To account for the quasi-condensate fluctuations, we use the method introduced in \cite{shapiro2,shapiro1}. First, we  evaluate 
numerically the density profile of the pure condensate by finding
a ground state of the stationary Gross-Pitaevskii equation 
\begin{equation}\label{gp}
  \mu\varphi(\x)=\left(-\frac{\hbar^2\nabla^2}{2m}+\frac{m \omega_r^2}{2}r^2+\frac{m \omega_z^2}{2}z^2+g|\varphi(\x)|^2 \right) \varphi(\x),
\end{equation}
where $g=\frac{4\pi\hbar^2 a}m$ and $\mu$ is the chemical potential. 

Next, we construct the quasi-condensate wave function $\psi(\x)$ by imprinting a phase $\phi(z)$ onto the pure condensate function, $\psi(\x)=|\varphi(\x)|e^{i\phi(z)}$, 
where 
\begin{equation}\label{phase}
  \phi(z)=\sum_{j=1}^\infty \sqrt{\frac{g\,\omega_r^2\,(j+2)(2j+3)}{4\pi\, z^{3}_{\rm tf}\,\omega_z^2\,\epsilon_j\,(j+1)}} P^{(1,1)}_j\left(\frac{z}{z^{\phantom{2}}_{\rm tf}}\right) \alpha_j.
\end{equation}
Here, $\epsilon_j = \frac{\hbar \omega_z}2 \sqrt{j(j+3)}$ is the energy of the low-lying axial excitations and $P^{(1,1)}_j$ is the Jacobi polynomial.
Also, $z^{\phantom{2}}_{\rm tf}$ is the axial density width given by Thomas-Fermi approximation.

\begin{figure}[htb]
  \centering
  \includegraphics[scale=0.3,clip]{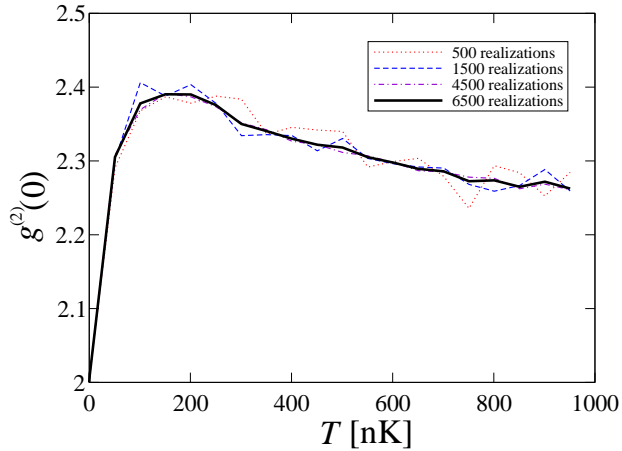}
  \caption{The peak of the second order correlation function $g^{(2)}(0)$ as a function of temperature. Different curves correspond to averaging over various number of realizations.
  Clearly, $g^{(2)}(0)\geq2$ and the inequality is saturated only for $T\simeq0$.}
  \label{fig_peak}
\end{figure}

The phase fluctuations result from randomness of $\alpha_j$, which is drawn from a Gaussian distribution with zero mean and the variance given by occupation of the $j$-th mode
\begin{equation}
  \langle\alpha_j^2\rangle=\frac1{\exp(\epsilon_j/k_B T) - 1}.
\end{equation}
Consequently, the temperature of the gas enters the dynamics of the Raman scattering. Due to presence of the $z$-dependent phase factor (\ref{phase}) in the quasi-condensate wave-function,
the momentum distribution along the $z$ axis broadens as the temperature grows.

The quasi-condensate is illuminated with an intense laser beam with the wave-vector
equal to $k_p = 5.80\times 10^{6}\, \f{m}^{-1}$. Since the Stokes photon and the pump wave-vectors are similar, we set $k_s \approx k_p$.

Our final simplification regards the form of the condensate density profile. A simple numerical check shows that the ground state of Eq.~(\ref{gp}) can be approximated by a Gaussian
function in the radial direction, so that
\begin{equation}\label{wave}
  \psi(\x)= \sqrt\frac{N\sigma_r^2}{\pi}e^{-\frac{\sigma_r^2}{2}(x^2+y^2)}\cdot \varphi(z)e^{i\phi(z)}.
\end{equation}
The axial function $\varphi(z)$ is found numerically by setting $x=y=0$ in Eq.~(\ref{gp}) and the Gaussian fit gives $\sigma_r\simeq0.10 k_s$. 
All the numerical results presented below are obtained by calculating the relevant physical quantity for a single realization of the phase noise 
$\phi(z)$ and then averaging over many such realizations.

We can now estimate the free expansion time $t_f$, at which the system enters the far-field regime in the $z$ direction. The velocity spread
of the quasi-condensate $\Delta k_z$ is approximately 2 mm/s, while the initial size is $2z_{\rm tf}\simeq 1$ mm. Therefore, the far-field condition would be
$t_f\gg 2z_{\rm tf}/\Delta k_z=0.5$ s.

Let us now comment on the consistency of the above approximations. Equation (\ref{phase}) was derived in \cite{shapiro1} under assumption, that the quasi-condensate has a Thomas-Fermi
density profile in all three dimensions, valid when the non-linear term dominates the Gross-Pitaevskii equation.
To simplify our calculations, we model radial wave-functions with Gaussians, which is true in the quasi-1D limit, when the radial trapping
potential dominates over the non-linear term. However, when we plot the numerically evaluated ground state of Eq.~(\ref{gp}), it turns out to be in an intermediate regime, and
could be equally well modeled with either a Gaussian or a Thomas-Fermi shape. Therefore, the above method can be regarded as an appropriate approach in such a transitional case. 
Since our main goal is to demonstrate the general behavior of the density and the correlation functions of scattered atoms as a function of $T$, we believe that this approximate method
is sufficiently precise for the purpose.

\subsection{Density of scattered atoms as a function of $T$}

First, we present the numerical results for the momentum distribution of scattered atoms, as given by Eq.~(\ref{dens}). 
We use the reference frame
co-moving with a velocity $\hbar\kk_p/m$, hence according to Fig.~\ref{fig:schematic_halo} the density is centered around $\kk=0$ with the radius equal to $k_s$. We investigate
the momentum density as a function of $k_z$ in a vicinity of $\kk=k_s{\bf e}_z$. This quantity, via Eq.~(\ref{dens}), 
samples the $z$-dependence of the momentum distribution of the quasi-condensate and therefore may provide some information on its temperature.

In Fig.~\ref{den_int} we schematically show which $k_s{\bf n}$ vectors contribute to the integral Eq.~(\ref{dens}) with $\kk\simeq k_s{\bf e}_z$. When $k_z=k_s$, 
as in the main part of the Figure, the integration runs approximately through the center of the cloud. When $k_z<k_s$, as in the inset (a), the tails of the quasi-condensate
still contribute to the density so the integral does not vanish rapidly as we move away from $k_z=k_s$. On the other hand, when $k_z>k_s$, shown in (b), the intergal does not
sample the tails anymore and the density of scattered atoms quickly drops with growing $k_z$.

This simple graphical interpretation is readily confirmed in Fig.~\ref{kz_den}, where we present the result of numerical integration of Eq.~(\ref{dens}) with the quasi-condensate function
obtained for various temperatures and averaged over 400 realizations of the phase noise. 
As expected, for $T=0.1\,$nK the density is peaked around $k_z\simeq k_s$, and is largely extended for $k_z<k_s$, while it drops immediately as $k_z>k_s$. 
As the temperature grows, the density widens substantially due to increased width of the quasi-condensate in the momentum space.
Therefore, the density of scattered atoms, when measured along the $k_z$ axis, could be used to determine the temperature of the mother cloud.

Next, we investigate the dependence of the atomic density measured in position space after a typical expansion time of $t_f = 300\,$ms \cite{paris1,paris2,paris3}. At this time,
as argued in the previous section, the system has not yet entered the far-field regime. 
We set $\frac{m{\bf r}}{\hbar t_f}\simeq k_s{\bf e_z}$ in Eq.~(\ref{dens_pos})
to make a direct comparison with previous results and evaluate the integral numerically. We observe that when the expansion time is finite, 
contrary
to the far-field regime considered above, the density is less sensitive to the temperature. Also, we notice that for high $T$ results in both cases are more similar. 
\begin{figure}[htb]
  \centering
  \includegraphics[width=0.450\textwidth,clip]{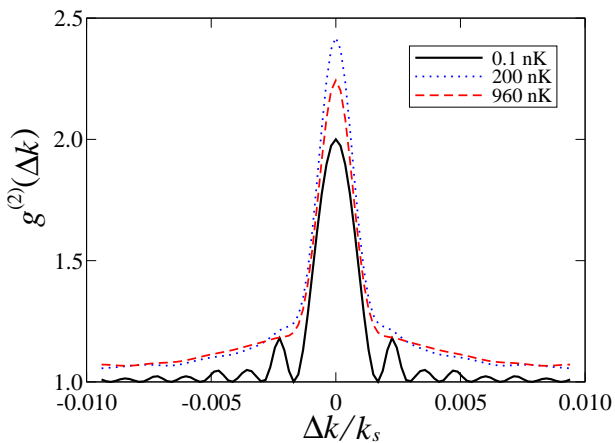}
  \caption{The dependence of the second order correlation function $g^{(2)}(\Delta k)$ on the length of the wave-vector $\Delta k$ in a vicinity of $k_s{\bf e}_z$.
    The black solid line is calculated for $T=0.1\,$nK, the dotted blue for $T=200\,$nK while the dashed red for $T=960\,$nK. The number of realizations was 400 for each curve.}
  \label{kz_corr}
\end{figure}
This is because at large temperatures, the momentum distribution of the quasi-condensate broadens and so the far-field condition is satisfied at earlier times.

\subsection{Dependence of correlation functions on the temperature}

We now investigate the impact of the phase fluctuations on the correlations of scattered atoms. We begin with the far-field expression (\ref{G1_momentum}) and using Eq.~(\ref{g2}) we 
calculate the normalized second order correlation function
\begin{equation}
  g^{(2)}(\kk_1,\kk_2)=\frac{\left\langle G^{(2)}(\kk_1,\kk_2)\right\rangle_{\phi}}{\left\langle n(\kk_1)\right\rangle_{\phi}\left\langle n(\kk_2)\right\rangle_{\phi}},
\end{equation}
where $\langle\cdot\rangle_{\phi}$ denotes averaging over many realizations of the phase $\phi(z)$. In order to be consistent with the results of the previous section, 
we concentrate on a region of wave-vectors close to $k_s{\bf e}_z$. Namely, we set $\kk_1=k_s{\bf e}_z+\frac{\Delta k}2{\bf e}_z$ and $\kk_2=k_s{\bf e}_z-\frac{\Delta k}2{\bf e}_z$
and analyze $g^{(2)}(\Delta k)$. First, note that for $\Delta k=0$, according to Eq.~(\ref{g2}) we have
$g^{(2)}(0)=\frac{2\langle n^2\rangle_{\phi}}{\langle n\rangle^2_{\phi}}$, where  $n=G^{(1)}(k_s{\bf e}_z,k_s{\bf e}_z)$. Since the variance 
$\langle n^2\rangle_{\phi}-\langle n\rangle^2_{\phi}$ is non-negative, then $g^{(2)}(0)\geq2$. The inequality is saturated only in the absence of noise fluctuations. To picture the impact
of the temperature on the height of the peak of the second order correlation function, in Fig.~\ref{fig_peak} we plot $g^{(2)}(0)$ for various temperatures. We clearly notice
the change of the height of the peak as soon as $T>0$.

\begin{figure}[htb]
  \centering
  \includegraphics[width=0.450\textwidth,clip]{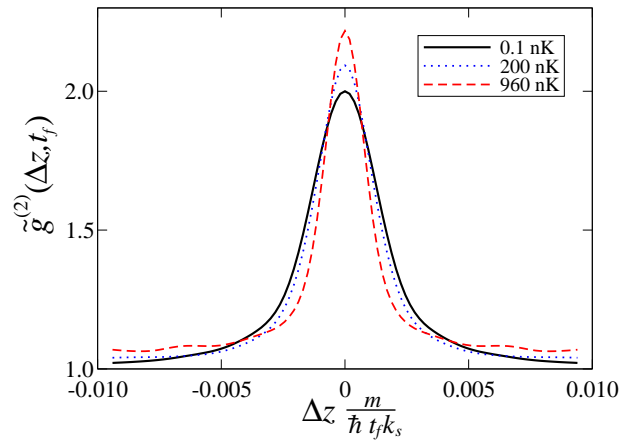}
  \caption{The dependence of the second order correlation function $\tilde g^{(2)}(\Delta z,t_f)$ in the position space in a vicinity of $\frac{m\x}{\hbar t_f}=k_s{\bf e}_z$
    as a function of $\Delta z$.
    The black solid line is calculated for $T=0.1\,$nK, the dotted blue for $T=200\,$nK while the dashed red for $T=960\,$nK. The number of realizations was 400 for each curve.}
  \label{z_corr}
\end{figure}

Next, in Fig.~\ref{kz_corr} we plot $g^{(2)}(\Delta k)$ as a function of $\Delta k$ for various temperatures averaged for 400 realizations. We observe that apart from the change of the
peak height $g^{(2)}(0)$, the wings of the correlation function broaden, due to increased momentum width of the quasi-condensate at higher temperatures. When $T\simeq0$, the correlation
function oscillates in the momentum space. This behavior is determined by a Fourier transform of the Thomas-Fermi profile in the $k_z$ direction. At larger temperatures, 
the phase fluctuations smear out the fringes.

We now switch to the finite expansion time regime and evaluate the normalized second order correlation function in position space. 
Namely, we use the definition (\ref{g2}), where the $G^{(1)}$ 
function is calculated using Eq.~(\ref{G1_position}), giving
\begin{equation}
  \tilde g^{(2)}(\x_1,\x_2,t_f)=\frac{\left\langle G^{(2)}(\x_1,\x_2,t_f)\right\rangle_{\phi}}{\left\langle n(\x_1,t_f)\right\rangle_{\phi}\left\langle n(\x_2,t_f)\right\rangle_{\phi}}.
\end{equation}

In analogy to the far-field case, we set 
${\bf r}_1=\frac{\hbar t_f}mk_s{\bf e}_z+\frac{\Delta z}2{\bf e}_z$ and ${\bf r}_2=\frac{\hbar t_f}mk_s{\bf e}_z-\frac{\Delta z}2{\bf e}_z$. In Fig.~\ref{z_corr} 
we plot $\tilde g^{(2)}(\Delta z,t_f)$ as a function of $\Delta z$ for three different values of temperature $T$ and $t_f=300$ ms. 
Equivalently to the results obtained in the momentum space, we have $\tilde g^{(2)}(0,t_f)\geq2$ and agian we observe that this equality is saturated only for $T\simeq0$. However,
differently from the previous case, the oscillations
of the correlation function at low temperatures are not present, since the Thomas-Fermi profile is a smooth function of $z$. Also, the broadeding of the correlation function is
much less pronounced.



\section{Conclusions}

We have analyzed in detail the properties of the field of atoms scattered out of a quasi-condensate in the Raman process. 
We have demonstrated that the density of scattered
atoms, when measured in the far-field regime, strongly depends on the temperature of the quasi-condensate. However, this dependence is much weaker, when the expansion time is finite.
Furthermore,
we have calculated the second order correlation function in both expansion time regimes. 
In each case, $g^{(2)}$ broadens with growing $T$, although in the latter the effect
is less pronounced. The presence of the temperature-induced phase fluctuations can be also deduced from the peak height of $g^{(2)}$. While for the pure condensate, $g^{(2)}(0)=2$, 
this value can substantially rise at higher $T$.

In summary, the measurements of the position of scattered atoms could provide some information on the temperature of the mother quasi-condensate. Nevertheless, physical quantities
such as the density or the correlation functions do not change drastically in a wide range of temperatures $T\in[0,1]\,\mu$K. 
If the experiment is aimed at determining the temperature of the quasi-condensate, it requires very high spatial resolution low detection noise and long expansion time.

Note that in our calculations we have neglected the possible impact of the atomic transition rules on the field of scattered atoms. In the case of particular atomic transitions, 
due to polarization of the Stokes field, some scattering directions are forbidden. 
We underline, that such effect could be easily taken into account by modifying the coupling function $h(\kk_1,\kk_2)$.


\section{Acknowledgments}

We aregrateful to Wojciech Wasilewski, Marie Bonneau, Denis Boiron and Chris Westbrook for fruitful discussions. 
T.W. acknowledges the Foundation for Polish Science International Ph.D. Projects Program co-financed by the EU European Regional Development Fund.
J. Ch. acknowledges Foundation for Polish Science International TEAM Program co-financed by the EU European Regional Development Fund. 
P.Z. acknowledges the support of Polish Ministry of Science and Higher
Education program ``Mobility Plus''. M.T. acknowledges financial support of the National Science Centre. 

\appendix\section{Perturbative solution for the atoms}
\label{app_time_dep_b}

In this Appendix we present details of the derivation of Eq.(\ref{b_time_dependence}). First, we introduce a solution of Eq.~(\ref{heis_a}) in absence of coupling, i.e.
\begin{equation}
  \hat a(\kk,t) = \hat a(\kk)e^{-i \omega_k t}.
\end{equation}
This expression, when inserted into Eq.~(\ref{heis_b}), gives a first order equation of motion for the atomic field, which reads
\begin{eqnarray}
  \partial_t \hat\beta(\kk,t) &=& \frac{1}{i \hbar}\int\!\!d\kk' h(\kk',\kk) \hat a^{\dagger}(\kk') e^{i(\omega_{k'}+\frac{\epsilon_k}\hbar)t},
\end{eqnarray}
where $\hat b(\kk,t) = \hat\beta(\kk,t) e^{- i \epsilon_k t/ \hbar}$. Integration over time gives
\begin{eqnarray}
  \hat\beta(\kk,t)&=&\hat\beta(\kk,0)+\frac{t}{i\hbar}\!\! \int\!\!d\kk\, h(\kk,\kk') a^\dagger(\kk')\times \nonumber\\
  &\times&\f{sinc}\left(\left((\omega_{k'}+\frac{\epsilon_k}{\hbar}\right)\frac t2\right)e^{i(\omega_{k'}+\frac{\epsilon_k}\hbar)\frac t2}.
\end{eqnarray}
The typical values of the kinetic energy of scattered atoms are $\epsilon_k\simeq\frac{\hbar^2k_p^2}{2m}$, while the photon energies are of the order of $\omega_{k'}\simeq\hbar k_pc$.
Since $m c/\hbar \sim 1.9\cdot10^{16} \f{m}^{-1}$ is of the order of the inverse of ``Compton wavelength'' of an atom
one can drop the dependence on the atom energy in the ``sinc'' function. We now express the above equation in terms of operator $\hat b$ and 
arrive at Eq.~(\ref{b_time_dependence}).

\section{$G^{(1)}$ in momentum space}
\label{app_G1}

In this  Appendix we derive Eq.(\ref{G1_momentum}). Using the perturbative solution from Eq.~(\ref{b_time_dependence}) and the definition of $G^{(1)}$ from Eq.~(\ref{g1}), 
we obtain up to an irrelevant phase factor
\begin{equation}
  \label{G1_general}
  G^{(1)}(\kk_1,\kk_2) = \frac{t_p^2}{\hbar^2} \int\!\!d\kk' h^\star(\kk_1,\kk') h(\kk_2,\kk')\f{sinc}^2\!\left(\frac{\omega_{k'}t_p}2\right).
\end{equation}
Typically the duration of the pump pulse is of the order of $5\,\mu$s. For such value, $\frac1{ct_p} \sim 1.5\cdot10^{-3}\,\frac1{\f{m}}$ 
is much smaller than the width of the coupling function $h$, which, via Eq.~(\ref{h_def_fourier}), is related to the Fourier transform of the quasi-condensate function. Since
$\omega_{k'}$ is centered around $\omega_s$, we can set $|\kk'| = k_s$ in the coupling function and perform the integral over $k'$. This way, we obtain
\begin{equation}\label{app_g1}
  G^{(1)}(\kk_1,\kk_2)\simeq\frac{2\pi t_pk_s^2}{c\hbar^2}\int\!\!d\Omega\ h^*(\kk_1,k_s \B{n}') h(\kk_2,k_s \B{n}').
\end{equation} 
where $\kk' = k_s\B{n}'$, $|\B{n}'|=1$ and  $\int\! d\Omega'$ denotes integration over a solid angle pointed by  $\B n'$. Using the definition of the coupling function 
from Eq.~(\ref{h_def_fourier}) we arrive at Eq.~(\ref{G1_momentum}).

\section{$G^{(1)}$ in position space}
\label{app_position_rep}

In this Appendix, we calculate the first order correlation function
 in the position space after $t_f$ time of the free expansion. We employ a reasonable approximation, that the pump duration time
$t_p$ is much shorter than $t_f$. In this case, the position-dependent correlation function $\tilde G(\kk_1,\kk_2,t_f)$ is simply given by the following 
Fourier transform of Eq.~(\ref{app_g1})
\begin{equation}
  \tilde G^{(1)}(\x_1,\x_2,t_f) = \alpha \int\!\!d\Omega\ \Psi^*(\x_1,{\bf n},t_f)\Psi(\x_2,{\bf n},t_f),
\end{equation}
where $\alpha=\frac{2\pi t_pk_s^2}{c\hbar^2}|E_0|^2|h_0|^2$ and
\begin{equation}
  \Psi(\x,{\bf n},t_f)=\int\!\!\opi{3}{d\kk}\ e^{-i\frac{\epsilon_{k}t_f}\hbar + i\kk \cdot \x}  \tilde\psi(\kk + k_s \B{n}).
\end{equation}
Changing the variables to $\kk=\frac{m\x}{\hbar t_f} + \delta\kk$ gives Eq.~(\ref{G1_position}), up to an irrelevant phase factor.

\end{document}